\documentclass{ifae}

\usepackage[centertags]{amsmath}
\allowdisplaybreaks[4]
\numberwithin{equation}{section}
\usepackage{amsbsy}
\usepackage{amsfonts}
\usepackage{amssymb}
\usepackage{bm}
\usepackage{mathrsfs}
\usepackage{url}
\usepackage[dvips]{hyperref}

\usepackage[square,comma,numbers,sort&compress,nonamebreak]{natbib}

\newcommand{\vet}[1]{\ensuremath{\hskip-1pt\vec{\hskip1pt#1}}}

\begin{document}

\begin{header}
  \title{Theory of Neutrino Oscillations}

  \begin{Authlist}
    Carlo Giunti\Iref{to}

  \Affiliation{to}{
INFN, Sezione di Torino, and Dipartimento di Fisica Teorica,
Universit\`a di Torino,
Via P. Giuria 1, I--10125 Torino, Italy,
giunti@to.infn.it
  }
  \end{Authlist}

  
  \begin{abstract}
We present a derivation of the flavor neutrino states
which describe neutrinos produced or detected in
charged-current weak-interaction processes,
including those operating in neutrino oscillation experiments.
We also present a covariant derivation of the probability of neutrino oscillations
which is consistent with the fact that flavor is Lorentz-invariant.
Finally,
we clarify the negative answers to three commonly asked questions:
``Do charged leptons oscillate?'';
``Is the standard phase wrong by a factor of 2?''
``Are flavor neutrinos described by Fock states?''.
  \end{abstract} 
  
\end{header}


\section{Introduction}
\label{s1}

The study of neutrino oscillations is a very important field
of contemporary experimental ant theoretical
research in high-energy physics.
The main reason is that
neutrino oscillations is a consequence of the existence of neutrino masses,
as was discovered by Bruno Pontecorvo in the late 50's
\cite{Pontecorvo:1957cp,Pontecorvo:1958qd,Pontecorvo:1968fh}.
However,
neutrinos are massless in the Glashow-Salam-Weinberg Standard Model
\cite{Glashow-SM-61,Weinberg-SM-67,Salam-SM-68}.
Hence,
neutrino oscillations represents an open window on the physics beyond
the Standard Model.

In view of such importance of neutrino oscillations,
it is necessary to have a correct and clear
understanding of all aspects of the theory of neutrino oscillations,
in order to have a correct interpretation of experimental results.
The theory of neutrino oscillations has been discussed in many papers
\cite{Eliezer:1976ja,Fritzsch:1976rz,Bilenky:1976cw,Bilenky:1976yj,Nussinov:1976uw,Kayser:1981ye,Kobzarev:1982ra,Giunti:1991ca,Giunti:1992cb,Giunti:1992sx,Rich-93,hep-ph/9305276,Kiers:1996zj,hep-ph/9603430,Campagne-97,hep-ph/9703241,hep-ph/9709494,Kiers-Weiss-PRD57-98,hep-ph/9711363,Burkhardt:hep-ph/9803365,Nauenberg-Correlated-99,hep-ph/9909332,Giunti:2000kw,Giunti:2001kj,Giunti:2002ee,Giunti:2002xg,Giunti:2003ax,Burkhardt:hep-ph/0302084,physics/0305122,hep-ph/0402217}
and reviewed in
\cite{Bilenky:1978nj,Bilenky:1987ty,Boehm:1992nn,CWKim-book,Zralek-oscillations-98,Bilenkii:2001yh,Beuthe:2001rc,Dolgov:2002wy}
(conference contributions
and
papers with other points of view
are listed in Ref.~\cite{Neutrino-Unbound}).

In this report we first review briefly the standard plane-wave theory
of neutrino oscillations in Section~\ref{s2}.
In Section~\ref{s3} we discuss the problem of the definition of flavor neutrino states
and in Section~\ref{s4}
we show that the flavor neutrino states are appropriate for the
description of neutrinos produced or detected in
charged-current weak-interaction processes.
In Section~\ref{s5} we present a covariant derivation of the probability of neutrino oscillations.
In Section~\ref{s6} we try to give clear answers to three questions
which are often asked.
Finally, we conclude in Section~\ref{s7}.

\section{Standard Plane-Wave Theory}
\label{s2}

Neutrino oscillations are a consequence of
neutrino mixing:
the left-handed flavor neutrino fields $\nu_{\alpha L}$,
with $\alpha=e,\mu,\tau$,
are unitary linear combinations of the massive neutrino fields\footnote{
For simplicity we consider the simplest case of three-neutrino mixing,
neglecting the possible existence of additional sterile neutrino fields
(see Refs.~\cite{BGG-review-98,Gonzalez-Garcia:2002dz,hep-ph/0310238,hep-ph/0405172}).
}
$\nu_{kL}$,
with $k=1,2,3$,
\begin{equation}
\nu_{\alpha L}
=
\sum_{k=1}^3 U_{\alpha k} \, \nu_{kL}
\qquad
(\alpha=e,\mu,\tau)
\,,
\label{001}
\end{equation}
where $U$ is the unitary mixing matrix.

In the standard plane-wave theory
of neutrino oscillations
\cite{Eliezer:1976ja,Fritzsch:1976rz,Bilenky:1976cw,Bilenky:1976yj,Bilenky:1978nj}
it is assumed that the neutrinos created or detected together with
a charged antilepton\footnote{
Since $\alpha$ is an index,
we use the notation
$ \ell_e = e $,
$ \ell_\mu = \mu $,
$ \ell_\tau = \tau $.
}
$\ell_{\alpha}^+$,
with $\alpha=e,\mu,\tau$,
are described by the flavor neutrino state
\begin{equation}
| \nu_{\alpha} \rangle
=
\sum_{k=1}^3 U_{\alpha k}^* \, | \nu_k \rangle
\,.
\label{002}
\end{equation}
Since the massive neutrino states
$| \nu_k \rangle$
have definite mass $m_k$ and definite energy $E_k$,
they evolve in time as plane waves:
\begin{equation}
i
\,
\frac{\partial}{\partial t}
\,
| \nu_k(t) \rangle
=
\mathscr{H}
\,
| \nu_k(t) \rangle
=
E_k
\,
| \nu_k(t) \rangle
\quad \Longrightarrow \quad
| \nu_k(t) \rangle
=
e^{- i E_k t}
\,
| \nu_k \rangle
\,,
\label{003}
\end{equation}
where $\mathscr{H}$ is the Hamiltonian operator
and
$| \nu_k \rangle = | \nu_k(t=0) \rangle$.
The consequent time evolution of the flavor neutrino state (\ref{002})
is given by
\begin{equation}
| \nu_{\alpha}(t) \rangle
=
\sum_{k=1}^3 U_{\alpha k}^* \, e^{- i E_k t}
\,
| \nu_k \rangle
=
\sum_{\beta=e,\mu,\tau}
\left(
\sum_{k=1}^3 U_{\alpha k}^* \, e^{- i E_k t} \, U_{\beta k}
\right)
| \nu_{\beta} \rangle
\,,
\label{004}
\end{equation}
which shows that if the mixing matrix $U$ is different from unity
(\textit{i.e.} if there is neutrino mixing),
for $t>0$ the state $| \nu_{\alpha}(t) \rangle$,
which has pure flavor $\alpha$ at $t=0$,
is a superposition of different flavors.
The quantity in round parentheses in Eq.~(\ref{004})
is the amplitude of $\nu_\alpha\to\nu_\beta$ transitions
at the time $t$ after $\nu_\alpha$ production,
whose squared absolute value gives
the probability of a $\nu_\alpha\to\nu_\beta$ transitions:
\begin{equation}
P_{\nu_\alpha\to\nu_\beta}(t)
=
|\langle \nu_{\beta} | \nu_{\alpha}(t) \rangle |^2
=
\left|
\sum_{k=1}^3 U_{\alpha k}^* \, e^{- i E_k t} \, U_{\beta k}
\right|^2
=
\sum_{k,j=1}^3 U_{\alpha k}^* \, U_{\beta k} \,  U_{\alpha j} \, U_{\beta j}^*
\, e^{- i ( E_k - E_j ) t}
\,.
\label{005}
\end{equation}
One can see that
$P_{\nu_\alpha\to\nu_\beta}(t)$
depends on the energy differences
$E_k - E_j$.
In the standard theory of neutrino oscillations
it is assumed that all massive neutrinos have the same momentum $\vet{p}$.
Since detectable neutrinos are ultrarelativistic\footnote{
It is known that neutrino masses
are smaller than about one eV
(see Refs.~\cite{Bilenky:2002aw,hep-ph/0310238}).
Since only neutrinos with energy larger than about 100 keV
can be detected (see the discussion in Ref.~\cite{Giunti:2002xg}),
in oscillation experiments neutrinos are always ultrarelativistic.
},
we have
\begin{equation}
E_k
=
\sqrt{\vet{p}^2+m_k^2}
\simeq
E
+
\frac{m_k^2}{2E}
\quad \Longrightarrow \quad
E_k - E_j
=
\frac{\Delta{m}^2_{kj}}{2E}
\,,
\label{006}
\end{equation}
where $E \equiv |\vet{p}|$
is the energy of a massless neutrino
(or, in other words,
the neutrino energy in the massless approximation),
and
$\Delta{m}^2_{kj} \equiv m_k^2 - m_j^2$.
In order to obtain a measurable flavor transition probability,
it is necessary to convert the time $t$,
which is not measured in neutrino oscillation experiments,
in the known source-detector distance $L$.
Considering ultrarelativistic neutrinos,
we have $t \simeq L$,
leading to the standard formula for the oscillation probability
\begin{equation}
P_{\nu_\alpha\to\nu_\beta}(L)
=
\sum_{k,j=1}^3 U_{\alpha k}^* \, U_{\beta k} \,  U_{\alpha j} \, U_{\beta j}^*
\,
\exp\!\left(
- i \,
\frac{\Delta{m}^2_{kj} \, L}{2 \, E}
\right)
\,,
\label{007}
\end{equation}
which is used in the analyses of the experimental data on
neutrino oscillations in vacuum.

Summarizing,
there are three main assumptions in the standard theory of neutrino oscillations:
\renewcommand{\labelenumi}{\theenumi}
\renewcommand{\theenumi}{(A\arabic{enumi})}
\begin{enumerate}
\item
\label{A1}
Neutrinos produced or detected in
charged-current weak interaction processes
are described by the
flavor states (\ref{002}).
\item
\label{A2}
The massive neutrino states
$|\nu_{k}\rangle$
in Eq.~(\ref{002})
have the same momentum.
\item
\label{A3}
The propagation time
is equal to the
distance $L$
traveled by the neutrino
between production and detection.
\end{enumerate}

\section{Flavor Neutrino States}
\label{s3}

The flavor neutrino states which
describe neutrinos produced or detected in
charged-current weak interaction processes
can be derived
in the framework of Quantum Field Theory
\cite{Bilenkii:2001yh,Giunti:2002xg,hep-ph/0402217}.
The final state $|f\rangle$
in an interaction process is given by
\begin{equation}
|f\rangle
=
\bm{S} \, |i\rangle
\,,
\label{008}
\end{equation}
where $|i\rangle$ is the initial state and $\bm{S}$
is the S-matrix operator.
In general,
the final state can be a superposition of orthogonal states,
\begin{equation}
|f\rangle
=
\sum_{k=1}^3 \, \mathcal{A}_k \, |f_k\rangle
\,,
\label{009}
\end{equation}
with
$\langle f_k | f_j \rangle = \delta_{kj}$.
From Eq.~(\ref{008}) it follows that the amplitudes $\mathcal{A}_k$
are given by
\begin{equation}
\mathcal{A}_k
=
\langle f_k | f \rangle
=
\langle f_k | \, \bm{S} \, |i\rangle
\,.
\label{010}
\end{equation}
Turning now to the case of neutrinos,
for definiteness
let us consider neutrino production in the general decay process
\begin{equation}
P_I \to P_F + \ell_{\alpha}^+ + \nu_{\alpha}
\,,
\label{011}
\end{equation}
where $P_I$ is the decaying particle,
$P_F$ is a decay product
(which can be absent,
for example in $ \pi^+ \to \mu^+ + \nu_\mu $),
and $\alpha=e,\mu,\tau$.
Other production processes,
as well as detection processes, can be treated in a similar way
with straightforward modifications to the formalism.
The final state of $P_I$ decay,
\begin{equation}
| f \rangle
=
\bm{S} \, |P_I\rangle
\,,
\label{012}
\end{equation}
is given by
\begin{equation}
| f \rangle
=
N_{\alpha} \, | \nu_{\alpha} , \ell_{\alpha}^+ , P_F \rangle
+
\ldots
\,,
\label{012a}
\end{equation}
where $N_{\alpha}$
is a normalization coefficient
and the dots represents other decay channels and
the undecayed initial state $P_I$ itself.
The state
$| \nu_{\alpha} , \ell_{\alpha}^+ , P_F \rangle$
can be decomposed in a superposition of orthogonal states corresponding to
different massive neutrinos:
\begin{equation}
| \nu_{\alpha} , \ell_{\alpha}^+ , P_F \rangle
=
\frac{1}{N_{\alpha}} \,
\sum_{k=1}^3
\mathcal{A}_{\alpha k} \, | \nu_k , \ell_{\alpha}^+ , P_F \rangle
\,,
\label{013}
\end{equation}
with the amplitudes $\mathcal{A}_{\alpha k}$
given by
\begin{equation}
\mathcal{A}_{\alpha k}
=
\langle \nu_k , \ell_{\alpha}^+ , P_F | f \rangle
=
\langle \nu_k , \ell_{\alpha}^+ , P_F | \, \bm{S} \, |P_I\rangle
\,.
\label{014}
\end{equation}
The normalization of the state (\ref{013})
requires that
\begin{equation}
|N_{\alpha}|^2
=
\sum_{k=1}^3
| \mathcal{A}_{\alpha k} |^2
\,.
\label{014a}
\end{equation}
It is convenient to write Eq.~(\ref{013}) in the form
\begin{equation}
| P_F , \ell_{\alpha}^+ \rangle \, | \nu_{\alpha} \rangle
=
| P_F , \ell_{\alpha}^+ \rangle
\left(
\sum_{k=1}^3
| \mathcal{A}_{\alpha k} |^2
\right)^{-1/2}
\sum_{k=1}^3
\mathcal{A}_{\alpha k} \, | \nu_k \rangle
\,,
\label{015}
\end{equation}
which shows that the final neutrino in the decay (\ref{011}) is described by the
flavor state
\begin{equation}
| \nu_{\alpha} \rangle
=
\left(
\sum_{k=1}^3
| \mathcal{A}_{\alpha k} |^2
\right)^{-1/2}
\sum_{k=1}^3
\mathcal{A}_{\alpha k} \, | \nu_k \rangle
\,,
\label{016}
\end{equation}
which is calculable through Eq.~(\ref{014}).
This flavor state has the same structure as the standard one in Eq.~(\ref{002}).
In order to see if the amplitudes
$
\displaystyle
\mathcal{A}_{\alpha k}
\left(
\sum_{k=1}^3
| \mathcal{A}_{\alpha k} |^2
\right)^{-1/2}
$
agree with the standard ones given by $U_{\alpha k}^*$,
we need to calculate the matrix elements in Eq.~(\ref{014}).
In the case of the low-energy weak decay (\ref{011})
the effective S-matrix operator can be approximated at first order
in the perturbative expansion as
\begin{equation}
\bm{S}
\simeq
\bm{1}
-
i \int \mathrm{d}^4x \, \mathscr{H}^{\mathrm{CC}}_{\mathrm{I}}(x)
\,,
\label{017}
\end{equation}
where $\mathscr{H}^{\mathrm{CC}}_{\mathrm{I}}(x)$
is the charged-current weak interaction Hamiltonian
\begin{align}
\mathscr{H}^{\mathrm{CC}}_{\mathrm{I}}(x)
=
\null & \null
\frac{G_{\mathrm{F}}}{\sqrt{2}}
\sum_{\alpha=e,\mu,\tau}
\overline{\nu_{\alpha}}(x)
\,
\gamma^{\rho}
\left( 1 - \gamma^5 \right)
\ell_{\alpha}(x)
\,
J_{\rho}^{P_I \to P_F}(x)
+
\text{H.c.}
\nonumber
\\
=
\null & \null
\frac{G_{\mathrm{F}}}{\sqrt{2}}
\sum_{\alpha=e,\mu,\tau}
\sum_{k=1}^3
U_{\alpha k}^*
\,
\overline{\nu_{k}}(x)
\,
\gamma^{\rho}
\left( 1 - \gamma^5 \right)
\ell_{\alpha}(x)
\,
J_{\rho}^{P_I \to P_F}(x)
+
\text{H.c.}
\,.
\label{018}
\end{align}
The operator
$J_{\rho}^{P_I \to P_F}(x)$
describes the transition $P_I \to P_F$.
For the matrix element
$\langle \nu_k , \ell_{\alpha}^+ , P_F | \, \bm{S} \, |P_I\rangle$
we obtain
\begin{equation}
\mathcal{A}_{\alpha k}
=
\langle \nu_k , \ell_{\alpha}^+ , P_F | \, \bm{S} \, |P_I\rangle
=
U_{\alpha k}^*
\,
\mathcal{M}_{\alpha k}
\,,
\label{019}
\end{equation}
with
\begin{equation}
\mathcal{M}_{\alpha k}
=
- i
\,
\frac{G_{\mathrm{F}}}{\sqrt{2}}
\int \mathrm{d}^4x \,
\langle \nu_k , \ell_{\alpha}^+ , P_F |
\,
\overline{\nu_{k}}(x)
\,
\gamma^{\rho}
\left( 1 - \gamma^5 \right)
\ell_{\alpha}(x)
\,
J_{\rho}^{P_I \to P_F}(x)
\,
| P_I \rangle
\,.
\label{020}
\end{equation}
Hence,
the flavor neutrino state (\ref{016})
is given by
\begin{equation}
| \nu_{\alpha} \rangle
=
\sum_{k=1}^3
\frac
{ \displaystyle \mathcal{M}_{\alpha k}}
{ \displaystyle \sqrt{ \sum_{j=1}^3 |U_{\alpha j}|^2 \, |\mathcal{M}_{\alpha j}|^2 } }
\,
U_{\alpha k}^*
\,
| \nu_{k} \rangle
\,.
\label{021}
\end{equation}
We notice that the flavor neutrino state (\ref{021})
has the structure of the standard flavor state (\ref{002}),
with the amplitudes of the massive neutrinos $ | \nu_k \rangle $
proportional to the corresponding element $U_{\alpha k}^*$
of the mixing matrix.
The additional coefficients
$
\displaystyle
\mathcal{M}_{\alpha k}
\left( \sum_{j=1}^3 |U_{\alpha j}|^2 \, |\mathcal{M}_{\alpha j}|^2 \right)^{-1/2}
$
disappear in the realistic case of ultrarelativistic neutrinos,
because in this case the experiments are not sensitive to the dependence
of $\mathcal{M}_{\alpha k}$ on the neutrino masses $m_k$,
leading to the approximation
\begin{equation}
\mathcal{M}_{\alpha k} \simeq \mathcal{M}_{\alpha}
\quad \Longrightarrow \quad
\frac
{ \displaystyle \mathcal{M}_{\alpha k}}
{ \displaystyle \sqrt{ \sum_{j=1}^3 |U_{\alpha j}|^2 \, |\mathcal{M}_{\alpha j}|^2 } }
\simeq
1
\,,
\label{022}
\end{equation}
where we have used the unitarity relation
$
\displaystyle
\sum_{j=1}^3 |U_{\alpha j}|^2 = 1
$.

Hence,
in the case of ultrarelativistic neutrinos we obtain the standard
flavor neutrino states (\ref{002})
in a rigorous way in the framework of Quantum Field Theory.
Since in all neutrino oscillation experiments the approximation (\ref{022})
is valid, in neutrino oscillations flavor neutrinos are correctly described
by the standard flavor states (\ref{002}),
confirming the validity of the standard
assumption~\ref{A1}.
Let us emphasize that the validity of this approximation is very important,
because
the standard flavor states (\ref{002})
do not depend on the kinematics of the process in which
the flavor neutrino is created or detected.
Hence,
it is possible to derive a general expression for the
oscillation probability,
as we will see in Section~\ref{s5}.

\section{Neutrino Production and Detection}
\label{s4}

In this Section we show that the flavor states (\ref{016})
are appropriate for the description of neutrinos produced or detected in
charged-current weak interaction processes.
We consider for simplicity only the decay process
(\ref{011}),
but the same reasoning can be applied with straightforward modifications
to any production or detection process.

It is well known
\cite{Shrock:1980vy,McKellar:1980cn,Kobzarev:1980nk,Shrock:1981ct,Shrock:1981wq}
that the probability of the decay (\ref{011})
is the sum of the decay probabilities in the different massive neutrinos
$\nu_k$
weighted by the squared absolute value $|U_{\alpha k}|^2$
of the element of the mixing matrix that weights the
contribution of $\nu_k$
to the charged-current weak interaction Hamiltonian (\ref{018}).
The reason is that the massive neutrinos
have definite kinematical properties and constitute the
possible orthogonal asymptotic states of the decay.
In other words,
each decay in a massive neutrino constitutes a possible
decay channel.
Hence,
the probability of the decay (\ref{011})
is given by
\begin{equation}
|\mathcal{A}_{\alpha}|^2
=
\sum_{k=1}^3 |U_{\alpha k}|^2 \, |\mathcal{M}_{\alpha k}|^2
\,,
\label{023}
\end{equation}
with $\mathcal{M}_{\alpha k}$ given by Eq.~(\ref{020}).

Let us now check that the description of the neutrino
$\nu_{\alpha}$ produced in the decay (\ref{011})
through the flavor state (\ref{016}) gives the correct result (\ref{023}).
Indeed,
the amplitude of the decay process (\ref{011})
is given by
\begin{equation}
\mathcal{A}_{\alpha}
=
\langle \nu_{\alpha} , \ell_{\alpha}^+ , P_F |
\,
\widehat{S}
\,
| P_I \rangle
=
\left( \sum_{k=1}^3 |\mathcal{A}_{\alpha k}|^2 \right)^{-1/2}
\sum_{k=1}^3 \mathcal{A}_{\alpha k}^*
\,
\langle \nu_{k} , \ell_{\alpha}^+ , P_F |
\bm{S}
| P_I \rangle
=
\left( \sum_{k=1}^3 |\mathcal{A}_{\alpha k}|^2 \right)^{1/2}
\,.
\label{1301}
\end{equation}
Therefore,
the decay probability is given by an incoherent
sum of the probabilities of production of different massive neutrinos:
using Eq.~(\ref{019}) we obtain
\begin{equation}
|\mathcal{A}_{\alpha}|^2
=
\sum_{k=1}^3 |\mathcal{A}_{\alpha k}|^2
=
|U_{\alpha k}|^2 \, |\mathcal{M}_{\alpha k}|^2
\,,
\label{1302}
\end{equation}
in agreement with Eq.~(\ref{023}).
Thus,
the coherent character of the flavor state (\ref{016})
is correctly irrelevant for the decay rate.

It is important to understand that this result is very important
for the interpretation of the data of neutrino oscillation experiments,
because in the analysis of these data
it is necessary not only to calculate the oscillation probability
but also the neutrino production and detection rates.
We have shown that both tasks can be accomplished in a consistent
framework with the description of
the produced and detected neutrinos through the flavor states
(\ref{016}).

\section{Covariant Plane Wave Neutrino Oscillations}
\label{s5}

Since
flavor is a Lorentz-invariant quantity
(for example, an electron is an electron in any reference frame),
the probability of neutrino oscillations
is Lorentz invariant
\cite{Giunti:2000kw,physics/0305122}.
Indeed,
the standard oscillation probability formula (\ref{007})
is Lorentz-invariant,
as shown in Ref.~\cite{physics/0305122}.
However,
the standard derivation of the oscillation probability
reviewed in Section~\ref{s2}
is not formulated in a covariant way.
In this Section we present a covariant derivation
of the oscillation probability which shows explicitly its Lorentz-invariance
\cite{Winter:1981kj,Giunti:2000kw,Bilenkii:2001yh,hep-ph/0311241,hep-ph/0401244,hep-ph/0402217}.

In the following derivation
we consider ultrarelativistic neutrinos.
Of course,
neutrinos are not ultrarelativistic in all reference frames\footnote{
We are grateful to Prof. D. V. Ahluwalia-Khalilova for asking by e-mail a stimulating question on this point.
}.
However,
since the oscillation probability
is Lorentz-invariant,
it can be evaluated in any frame
and we can choose to evaluate it in a frame where the massive neutrinos are ultrarelativistic.
Let us emphasize that this is not a peculiar choice,
because there is a continuous infinite set of reference frames
in which massive neutrinos are ultrarelativistic,
which includes the one in which the detector is at rest.
Indeed,
we will always work with explicitly Lorentz-invariant expressions for the oscillation probability,
because these expressions must be valid
in the continuous infinite set of reference frames
in which massive neutrinos are ultrarelativistic.

The amplitude of $ \nu_\alpha \to \nu_\beta $
transitions
at a distance $L$ and after a time $T$ from the production of the flavor neutrino $\nu_{\alpha}$
is given by
\begin{equation}
A_{\alpha\beta}(L,T)
=
\langle \nu_{\beta} |
e^{ -i \bm{E} T + i \bm{P} L }
| \nu_{\alpha} \rangle
\,,
\label{13052}
\end{equation}
where
$\bm{E}$ and $\bm{P}$
are, respectively, the energy and momentum operators.
Since the massive neutrinos have definite masses and kinematical properties,
using the flavor states (\ref{002}),
which are valid for ultrarelativistic neutrinos,
we obtain
\begin{equation}
A_{\alpha\beta}(L,T)
=
\sum_{k=1}^3
U_{\alpha k}^*
\,
U_{\beta k}
\,
e^{- i E_k T + i p_k L}
\,,
\label{1305}
\end{equation}
with
\begin{equation}
E_k = \sqrt{ p_k^2 + m_k^2 }
\,.
\label{13051}
\end{equation}
The oscillation amplitude (\ref{1305}) is explicitly Lorentz-invariant.
Let us notice that we do not adopt the standard assumption~\ref{A2}
of equal momentum for different massive neutrinos.
Indeed,
it can be shown that
in general massive neutrinos do not have neither equal momenta nor equal energies
\cite{Giunti:2000kw,Giunti:2001kj,Giunti:2003ax,Giunti:2003ax}.
However,
in spite of the unrealistic character of the standard assumption~\ref{A2}
we will see that the standard expression for the oscillation probability
is correct.

In oscillation experiments the neutrino propagation time $T$ is not measured.
As in the standard plane-wave theory
reviewed in Section~\ref{s2},
for ultrarelativistic neutrinos it is possible to approximate $T \simeq L$
(assumption~\ref{A3}) in the phase in Eq.~(\ref{1305}).
The reason is that
in reality
neutrinos are described by wave packets
\cite{Nussinov:1976uw,Kayser:1981ye,Giunti:1991ca,Giunti:1992sx,hep-ph/9305276,Kiers:1996zj,hep-ph/9709494,Kiers-Weiss-PRD57-98,hep-ph/9711363,hep-ph/9909332,Beuthe:2001rc,Giunti:2002ee,Giunti:2002xg,Giunti:2003ax,hep-ph/0402217},
which are localized on the production process
at the production time and
propagate between the production and detection processes with a group velocity
which is close to the velocity of light.
If the massive neutrinos are ultrarelativistic
and contribute coherently to the detection process,
their wave packets
overlap with the detection process
for an interval of time $[ T - \Delta T \,,\, T + \Delta T ]$,
with
\begin{equation}
T
=
\frac{L}{\overline{v}}
\simeq
L \left( 1 + \frac{\overline{m^2}}{2E^2} \right)
\,,
\qquad
\Delta T \sim \sigma_x
\,,
\label{505}
\end{equation}
where $\overline{v}$
is the average group velocity,
$\overline{m^2}$ is the average of the squared neutrino masses,
and
$\sigma_x$
is given by the spatial uncertainties of the production and detection processes
summed in quadrature
\cite{hep-ph/9711363}
(the spatial uncertainty of the production process
determines the size of the massive neutrino wave packets).
The correction $ L \overline{m^2} / 2E^2 $ to $T=L$
in Eq.~(\ref{505})
can be neglected,
because it gives corrections to the oscillation phases
which are of higher order in the very small ratios
$ m_k^2 / E^2 $.
The corrections due to
$\Delta T \sim \sigma_x$
are also negligible,
because in all realistic experiments
$ \sigma_x $
is much smaller than the oscillation length
$L^{\mathrm{osc}}_{kj} = 4 \pi E / \Delta{m}^2_{kj}$,
otherwise oscillations could not be observed
\cite{Kayser:1981ye,Giunti:1991ca,Beuthe:2001rc,Giunti:2003ax}.
One can summarize these arguments by saying that
the approximation $T \simeq L$ is correct
because
the phase of the oscillations
is practically constant over the interval of time in which the
massive neutrino wave packets overlap with the detection process.

Using the approximation
$T \simeq L$
the phase in Eq.~(\ref{1305}) becomes
\begin{equation}
- E_k T + p_k L
\simeq
- \left( E_k - p_k \right) L
=
- \frac{ E_k^2 - p_k^2 }{ E_k + p_k } \, L
=
- \frac{ m_k^2 }{ E_k + p_k } \, L
\simeq
- \frac{ m_k^2 }{ 2 E } \, L
\,,
\label{1005}
\end{equation}
where $E$ is the neutrino energy neglecting mass contributions.
Let us emphasize that both the left and right hand sides of
Eq.~(\ref{1005})
are Lorentz-invariant
\cite{physics/0305122}.

Inserting the expression (\ref{1005}) in Eq.~(\ref{1305})
one can obtain in a straightforward way
a probability of
$ \nu_\alpha \to \nu_\beta $
transitions in space which coincides with the standard one in Eq.~(\ref{007}).
Therefore,
the result of the covariant derivation of the oscillation probability
confirms the correctness of the standard formula in Eq.~(\ref{007}).

It is important to notice that Eq.~(\ref{1005})
shows that the phases of massive neutrinos relevant for the oscillations
are independent from the particular values of the energies and momenta
of different massive neutrinos
\cite{Winter:1981kj,Giunti:1991ca,Giunti:2000kw,Giunti:2001kj,Giunti:2003ax},
because of the relativistic dispersion relation (\ref{13051}).
Hence,
as remarked above,
the standard assumption~\ref{A2} of equal momenta for different massive neutrinos,
albeit unrealistic,
is irrelevant for the correctness of the oscillation probability.

\section{Questions}
\label{s6}

In this section we try to clarify the negative answers to three
questions which are often asked.

\subsection{Do Charged Leptons Oscillate? No!}
\label{s61}

In order to understand the negative answer to this question
it is important to comprehend that
the only characteristic which distinguishes different charged leptons is
their mass.
Thus,
flavor is distinguished through the mass of the corresponding
charged lepton
or,
in other words,
flavor and mass coincide for charged leptons.
Since for charged leptons there is no mismatch between flavor and mass,
there cannot be flavor oscillations
of charged leptons
\cite{Giunti:2000kw}.

In Refs.~\cite{hep-ph/9509261,hep-ph/9707268,Srivastava:1998gi}
it has been claimed that the probability to detect a charged lepton
oscillates in space-time.
This claim has been refuted in
Ref.~\cite{hep-ph/9703241}.
A similar effect,
called ``Lambda oscillations'',
which has been claimed to exist
\cite{Srivastava:1995ws,hep-ph/9605399}
for the $\Lambda$'s
produced together with a neutral kaon,
as in the process
$ \pi^- + p \to \Lambda + K^0 $,
has been refuted in
Refs.~\cite{Lowe:hep-ph/9605234,Burkhardt:hep-ph/9803365}.
Considering the pion decay process
$ \pi^- \to \mu^- + \bar\nu_\mu $,
the authors of Refs.~\cite{hep-ph/9509261,hep-ph/9707268,Srivastava:1998gi,Srivastava:hep-ph/9807543}
argued that, since the final state of the muon and antineutrino is entangled,
if the probability to detect the antineutrino oscillates as a function of distance,
also the probability to detect the muon must oscillate.
However,
it is well known that
the probability to detect the antineutrino,
irrespective of its flavor,
does not oscillate.
This property is usually called ``conservation of probability''
or ``unitarity''
and is represented mathematically by the general relation
$
\sum_\beta
P_{\bar\nu_\mu\to\bar\nu_\beta}
=
1
$.
Hence,
this argument actually proves that the probability to detect a charged lepton
does \emph{not} oscillates in space-time!

\subsection{Is the Standard Phase Wrong by a Factor of 2? No!}
\label{s62}

It as been claimed
that
the standard phase of neutrino oscillations in Eq.~(\ref{007}),
\begin{equation}
\Phi_{kj}
=
- \frac{\Delta{m}^2_{kj} \, L }{ 2 \, E }
\,,
\label{025}
\end{equation}
is wrong by a factor of two
\cite{Field:hep-ph/0211199}
or there is an ambiguity by a factor of two in the oscillation phase
\cite{hep-ph/9607201,Lipkin:hep-ph/9901399,hep-ph/9906460,DeLeo:hep-ph/0208086}.
A similar discrepancy by a factor of two in the phase of $K^0-\bar{K}^0$
oscillations has been claimed in Refs.~\cite{Srivastava:1995bg,Srivastava:1995ws,Widom:hep-ph/9605399}
and an ambiguity by a factor of two has been claimed in Ref.~\cite{hep-ph/9501269}.
These claims,
which have been refuted in
Refs.~\cite{Nieto:hep-ph/9509370,Kayser:1995bw,Lowe:hep-ph/9605234,Kayser:1997fr,Kayser:hep-ph/9702327,Giunti:2000kw,Giunti:2002ee,Burkhardt:hep-ph/0302084},
stem from the following fallacious reasoning.

Different massive neutrinos
propagate with different velocities
\begin{equation}
v_k = \frac{p_k}{E_k}
\,,
\label{h160}
\end{equation}
where $E_k$ and $p_k$
are, respectively, the energy and momentum of the neutrino with mass $m_k$,
related by the relativistic dispersion relation (\ref{13051}).
According to the fallacious reasoning,
the phases of the different massive neutrinos
wave functions after a propagation distance $L$
should take into account the different times of propagation of different massive neutrinos:
\begin{equation}
\tilde\Phi_k
=
p_k \, L
- E_k \, t_k
\,.
\label{h161}
\end{equation}
The propagation times are given by
\begin{equation}
t_k
=
\frac{L}{v_k}
=
\frac{E_k}{p_k} \, L
\,,
\label{h162}
\end{equation}
which lead, in the relativistic approximation, to the phase difference
\begin{equation}
\Delta\tilde\Phi_{kj}
\equiv
\tilde\Phi_k - \tilde\Phi_j
=
- \frac{\Delta{m}^2_{kj} \, L }{ E }
\,.
\label{h163}
\end{equation}
This phase difference is twice of the standard one in Eq.~(\ref{025}).

Let us notice that in Eq.~(\ref{h161})
we have considered the possibility of different energies and momenta
for different massive neutrino wave functions,
as we have done in the covariant derivation
of the neutrino oscillation probability
in Section~\ref{s5}.
The authors of Refs.~\cite{hep-ph/9501269,hep-ph/9607201,Lipkin:hep-ph/9901399}
claimed that a correct way to obtain the standard oscillation phase
is to assume the same energy for the different massive neutrino wave functions.
Apart from the fact that this is an unphysical assumption
\cite{Giunti:2000kw,Giunti:2001kj,Giunti:2003ax,Giunti:2003ax},
it is not true that the disagreement of a factor of two disappears
assuming the same energy for the different massive neutrino wave functions,
as clearly shown by the above derivation of Eq.~(\ref{h163}),
in which the energies of the different massive neutrino wave functions
could have been taken to be equal
\cite{Rotelli-99,Giunti:2002ee}.
Indeed,
even if the different massive neutrino wave functions
have the same energy,
the time contribution
$-E t_k+E t_j$
to the phase difference $\Delta\tilde\Phi_{kj}$
does not disappear if
$t_k \neq t_j$.
This contribution has been missed in
Refs.~\cite{hep-ph/9501269,hep-ph/9607201,Lipkin:hep-ph/9901399}.

The mistake in the fallacious reasoning described above
is due to a wrong use of the group velocity (\ref{h160})
in the phase (\ref{h161}),
which cannot depend on the group velocity.
The group velocity (\ref{h160})
is the velocity of the factor which modulates the amplitude
of the wave packet of the corresponding massive neutrino.
In neutrino oscillation experiments
the envelopes of the wave packets of the different massive neutrinos
propagate with the corresponding group velocity
and take different times
to cover the distance between the source and the detector.
However,
these different propagation times have
no effect on the phases of the wave functions
of the different massive neutrinos,
whose interference generates the oscillations.
Since the interference is a local effect,
it must be calculated at the same time,
as well as in the same point,
for all the massive neutrino contributions
\cite{Nieto:hep-ph/9509370,Kayser:1995bw,Lowe:hep-ph/9605234,Kayser:1997fr,Kayser:hep-ph/9702327,Giunti:2000kw,Giunti:2002ee,Burkhardt:hep-ph/0302084},
as we have done in Section~\ref{s5}.
Since
only the amplitude of the wave function of each massive neutrino is
determined by the wave packet envelope,
the only possible effect of
the different arrival times of the envelopes of the wave packets
of different massive neutrinos
is a reduction of the overlap of the different wave packets,
leading to a decoherence effect
\cite{Nussinov:1976uw,Kayser:1981ye,Giunti:1991ca,Giunti:1992sx,hep-ph/9305276,Kiers:1996zj,hep-ph/9709494,Kiers-Weiss-PRD57-98,hep-ph/9711363,hep-ph/9909332,Beuthe:2001rc,Giunti:2002ee,Giunti:2002xg,Giunti:2003ax,hep-ph/0402217}.
For more details, see Ref.~\cite{Giunti:2002ee}.

\subsection{Are Flavor Neutrinos Described by Fock States? No!}
\label{s63}

It must be noted that the flavor state (\ref{002}) is \emph{not}
a quantum of the flavor field $\nu_{\alpha}$
\cite{Giunti:1992cb}.
Indeed,
one can easily check that
the flavor state (\ref{002}) is not annihilated by the flavor field $\nu_{\alpha}$
if the neutrino masses are taken into account.
It is, however,
possible to construct a Fock space for the quantized flavor fields
\cite{Blasone:1995zc,Blasone:1998hf,Fujii:1998xa,hep-ph/9907382,Fujii:2001zv,Blasone:2002jv,Blasone:2002wp}.
Then,
it is natural to ask if the flavor Fock states describe
real neutrinos,
produced and detected in charged-current weak interaction processes.
Let us notice that,
since the flavor Fock states are different from the standard flavor states (\ref{002})
and the ``exact''
flavor states (\ref{016}) that we have derived in the framework of
Quantum Field Theory,
if the answer to the question under discussion were positive,
the theory of neutrino oscillations
would have to be revised,
as claimed in Refs.~\cite{Blasone:1995zc,Blasone:1998hf,Fujii:1998xa,hep-ph/9907382,Fujii:2001zv,Blasone:2002jv,Blasone:2002wp}.
However,
the answer is negative,
as explained in Ref.~\cite{hep-ph/0312256},
to which we direct the interested reader
for the detailed proof.
Here we mention only that the unphysical character of
the flavor Fock states can be proved by
\textit{reductio ad absurdum}:
the description of neutrinos created or detected in
charged-current weak interaction processes
through the flavor Fock states
leads to unphysical results.
Let us emphasize that
this fact precludes the description of neutrinos in oscillation experiments
through the flavor Fock states,
because
these neutrinos must be produced and detected in
weak interaction processes.
Hence,
the flavor Fock states
are ingenious mathematical constructs without
relevance for the description of real neutrinos.

\section{Conclusions}
\label{s7}

We have shown that the standard flavor states (\ref{002}) used in the derivation of
neutrino oscillations
can be derived in the framework of Quantum Field Theory
in the realistic case of ultrarelativistic neutrinos (Section~\ref{s3}).
In Section~\ref{s4} we have shown that
the ``exact'' flavor states (\ref{016})
are appropriate for the description of neutrinos produced or detected in
charged-current weak-interaction processes,
taking into account the neutrino masses.
In Section~\ref{s5}
we have presented a covariant derivation of the probability of neutrino oscillations
which is consistent with the fact that flavor is Lorentz-invariant.
Finally,
in Section~\ref{s6} we have discussed the negative answers to three questions
which are often asked.
In conclusion,
we would like to emphasize that
the standard expression (\ref{007}) for the probability
of neutrino oscillations in vacuum
has been proved to be correct in the framework of Quantum Field Theory
and all contrary claims stem from some misunderstanding.


\begin{thebibliography}{86}
\expandafter\ifx\csname natexlab\endcsname\relax\def\natexlab#1{#1}\fi
\expandafter\ifx\csname url\endcsname\relax
  \def\url#1{{\tt #1}}\fi

\bibitem[Pontecorvo(1957)]{Pontecorvo:1957cp}
B.~Pontecorvo,
\newblock ``{Mesonium and antimesonium}'',
\newblock {\em Sov. Phys. JETP}, 6, 429, 1957,
\newblock [Zh. Eksp. Teor. Fiz. 33, 549 (1957)].

\bibitem[Pontecorvo(1958)]{Pontecorvo:1958qd}
B.~Pontecorvo,
\newblock ``{Inverse beta processes and nonconservation of lepton charge}'',
\newblock {\em Sov. Phys. JETP}, 7, 172--173, 1958,
\newblock [Zh. Eksp. Teor. Fiz. 34, 247 (1958)].

\bibitem[Pontecorvo(1968)]{Pontecorvo:1968fh}
B.~Pontecorvo,
\newblock ``{Neutrino experiments and the question of leptonic-charge
  conservation}'',
\newblock {\em Sov. Phys. JETP}, 26, 984--988, 1968.

\bibitem[Glashow(1961)]{Glashow-SM-61}
S.~L. Glashow,
\newblock ``{Partial symmetries of weak interactions}'',
\newblock {\em Nucl. Phys.}, 22, 579--588, 1961.

\bibitem[Weinberg(1967)]{Weinberg-SM-67}
S.~Weinberg,
\newblock ``{A model of leptons}'',
\newblock {\em Phys. Rev. Lett.}, 19, 1264--1266, 1967.

\bibitem[Salam(1969)]{Salam-SM-68}
A.~Salam,
\newblock ``{Weak and electromagnetic interactions}'',
\newblock 1969,
\newblock Proc. of the 8$^{\mathrm{th}}$ Nobel Symposium on \textit{Elementary
  particle theory, relativistic groups and analyticity}, Stockholm, Sweden,
  1968, edited by N. Svartholm, p.367-377.

\bibitem[Eliezer and Swift(1976)]{Eliezer:1976ja}
S.~Eliezer and A.~R. Swift,
\newblock ``{Experimental consequences of electron neutrino - muon neutrino
  mixing in neutrino beams}'',
\newblock {\em Nucl. Phys.}, B105, 45, 1976.

\bibitem[Fritzsch and Minkowski(1976)]{Fritzsch:1976rz}
H.~Fritzsch and P.~Minkowski,
\newblock ``{Vector - like weak currents, massive neutrinos, and neutrino beam
  oscillations}'',
\newblock {\em Phys. Lett.}, B62, 72, 1976.

\bibitem[Bilenky and Pontecorvo(1976{\natexlab{a}})]{Bilenky:1976cw}
S.~M. Bilenky and B.~Pontecorvo,
\newblock ``{The lepton-quark analogy and muonic charge}'',
\newblock {\em Sov. J. Nucl. Phys.}, 24, 316--319, 1976{\natexlab{a}},
\newblock [Yad. Fiz. 24 (1976) 603]. URL:
  \href{http://www.nu.to.infn.it/pap/bilenky/BP-YF24-603-1976.pdf}{\url{http://www.nu.to.infn.it/pap/bilenky/BP-YF24-603-1976.pdf}}.

\bibitem[Bilenky and Pontecorvo(1976{\natexlab{b}})]{Bilenky:1976yj}
S.~M. Bilenky and B.~Pontecorvo,
\newblock ``{Again on neutrino oscillations}'',
\newblock {\em Nuovo Cim. Lett.}, 17, 569, 1976{\natexlab{b}}.

\bibitem[Nussinov(1976)]{Nussinov:1976uw}
S.~Nussinov,
\newblock ``{Solar neutrinos and neutrino mixing}'',
\newblock {\em Phys. Lett.}, B63, 201--203, 1976.

\bibitem[Kayser(1981)]{Kayser:1981ye}
B.~Kayser,
\newblock ``{On the quantum mechanics of neutrino oscillation}'',
\newblock {\em Phys. Rev.}, D24, 110, 1981.

\bibitem[Kobzarev et~al.(1982)Kobzarev, Martemyanov, Okun, and
  Shchepkin]{Kobzarev:1982ra}
I.~Y. Kobzarev, B.~V. Martemyanov, L.~B. Okun, and M.~G. Shchepkin,
\newblock ``{Sum rules for neutrino oscillations}'',
\newblock {\em Sov. J. Nucl. Phys.}, 35, 708, 1982.

\bibitem[Giunti et~al.(1991)Giunti, Kim, and Lee]{Giunti:1991ca}
C.~Giunti, C.~W. Kim, and U.~W. Lee,
\newblock ``{When do neutrinos really oscillate?: Quantum mechanics of neutrino
  oscillations}'',
\newblock {\em Phys. Rev.}, D44, 3635--3640, 1991.

\bibitem[Giunti et~al.(1992{\natexlab{a}})Giunti, Kim, and Lee]{Giunti:1992cb}
C.~Giunti, C.~W. Kim, and U.~W. Lee,
\newblock ``{Remarks on the weak states of neutrinos}'',
\newblock {\em Phys. Rev.}, D45, 2414--2420, 1992{\natexlab{a}}.

\bibitem[Giunti et~al.(1992{\natexlab{b}})Giunti, Kim, and Lee]{Giunti:1992sx}
C.~Giunti, C.~W. Kim, and U.~W. Lee,
\newblock ``{Coherence of neutrino oscillations in vacuum and matter in the
  wave packet treatment}'',
\newblock {\em Phys. Lett.}, B274, 87--94, 1992{\natexlab{b}}.

\bibitem[Rich(1993)]{Rich-93}
J.~Rich,
\newblock ``{The Quantum mechanics of neutrino oscillations}'',
\newblock {\em Phys. Rev.}, D48, 4318--4325, 1993.

\bibitem[Giunti et~al.(1993)Giunti, Kim, Lee, and Lee]{hep-ph/9305276}
C.~Giunti, C.~W. Kim, J.~A. Lee, and U.~W. Lee,
\newblock ``{Treatment of neutrino oscillations without resort to weak
  eigenstates}'',
\newblock {\em Phys. Rev.}, D48, 4310--4317, 1993,
\newblock
  \texttt{\href{http://arxiv.org/abs/hep-ph/9305276}{\url{hep-ph/9305276}}}.

\bibitem[Kiers et~al.(1996)Kiers, Nussinov, and Weiss]{Kiers:1996zj}
K.~Kiers, S.~Nussinov, and N.~Weiss,
\newblock ``{Coherence effects in neutrino oscillations}'',
\newblock {\em Phys. Rev.}, D53, 537--547, 1996,
\newblock
  \texttt{\href{http://arxiv.org/abs/hep-ph/9506271}{\url{hep-ph/9506271}}}.

\bibitem[Grimus and Stockinger(1996)]{hep-ph/9603430}
W.~Grimus and P.~Stockinger,
\newblock ``{Real Oscillations of Virtual Neutrinos}'',
\newblock {\em Phys. Rev.}, D54, 3414--3419, 1996,
\newblock
  \texttt{\href{http://arxiv.org/abs/hep-ph/9603430}{\url{hep-ph/9603430}}}.

\bibitem[Campagne(1997)]{Campagne-97}
J.~E. Campagne,
\newblock ``{Neutrino oscillations from pion decay in flight}'',
\newblock {\em Phys. Lett.}, B400, 135--144, 1997.

\bibitem[Dolgov et~al.(1997)Dolgov, Morozov, Okun, and
  Shchepkin]{hep-ph/9703241}
A.~D. Dolgov, A.~Y. Morozov, L.~B. Okun, and M.~G. Shchepkin,
\newblock ``{Do muons oscillate?}'',
\newblock {\em Nucl. Phys.}, B502, 3--18, 1997,
\newblock
  \texttt{\href{http://arxiv.org/abs/hep-ph/9703241}{\url{hep-ph/9703241}}}.

\bibitem[Giunti et~al.(1998)Giunti, Kim, and Lee]{hep-ph/9709494}
C.~Giunti, C.~W. Kim, and U.~W. Lee,
\newblock ``{When do neutrinos cease to oscillate?}'',
\newblock {\em Phys. Lett.}, B421, 237--244, 1998,
\newblock
  \texttt{\href{http://arxiv.org/abs/hep-ph/9709494}{\url{hep-ph/9709494}}}.

\bibitem[Kiers and Weiss(1998)]{Kiers-Weiss-PRD57-98}
K.~Kiers and N.~Weiss,
\newblock ``{Neutrino oscillations in a model with a source and detector}'',
\newblock {\em Phys. Rev.}, D57, 3091--3105, 1998,
\newblock
  \texttt{\href{http://arxiv.org/abs/hep-ph/9710289}{\url{hep-ph/9710289}}}.

\bibitem[Giunti and Kim(1998)]{hep-ph/9711363}
C.~Giunti and C.~W. Kim,
\newblock ``{Coherence of neutrino oscillations in the wave packet approach}'',
\newblock {\em Phys. Rev.}, D58, 017301, 1998,
\newblock
  \texttt{\href{http://arxiv.org/abs/hep-ph/9711363}{\url{hep-ph/9711363}}}.

\bibitem[Burkhardt et~al.(1999)Burkhardt, Lowe, Stephenson, and
  Goldman]{Burkhardt:hep-ph/9803365}
H.~Burkhardt, J.~Lowe, G.~J. Stephenson, and T.~Goldman,
\newblock ``{Oscillations of recoil particles against mixed states}'',
\newblock {\em Phys. Rev.}, D59, 054018, 1999,
\newblock
  \texttt{\href{http://arxiv.org/abs/hep-ph/9803365}{\url{hep-ph/9803365}}}.

\bibitem[Nauenberg(1999)]{Nauenberg-Correlated-99}
M.~Nauenberg,
\newblock ``{Correlated wave packet treatment of neutrino and neutral meson
  oscillations}'',
\newblock {\em Phys. Lett.}, B447, 23--30, 1999,
\newblock
  \texttt{\href{http://arxiv.org/abs/hep-ph/9812441}{\url{hep-ph/9812441}}}.

\bibitem[Cardall(2000)]{hep-ph/9909332}
C.~Y. Cardall,
\newblock ``{Coherence of neutrino flavor mixing in quantum field theory}'',
\newblock {\em Phys. Rev.}, D61, 073006, 2000,
\newblock
  \texttt{\href{http://arxiv.org/abs/hep-ph/9909332}{\url{hep-ph/9909332}}}.

\bibitem[Giunti and Kim(2001)]{Giunti:2000kw}
C.~Giunti and C.~W. Kim,
\newblock ``{Quantum mechanics of neutrino oscillations}'',
\newblock {\em Found. Phys. Lett.}, 14, 213--229, 2001,
\newblock
  \texttt{\href{http://arxiv.org/abs/hep-ph/0011074}{\url{hep-ph/0011074}}}.

\bibitem[Giunti(2001)]{Giunti:2001kj}
C.~Giunti,
\newblock ``{Energy and momentum of oscillating neutrinos}'',
\newblock {\em Mod. Phys. Lett.}, A16, 2363, 2001,
\newblock
  \texttt{\href{http://arxiv.org/abs/hep-ph/0104148}{\url{hep-ph/0104148}}}.

\bibitem[Giunti(2003{\natexlab{a}})]{Giunti:2002ee}
C.~Giunti,
\newblock ``{The phase of neutrino oscillations}'',
\newblock {\em Physica Scripta}, 67, 29--33, 2003{\natexlab{a}},
\newblock
  \texttt{\href{http://arxiv.org/abs/hep-ph/0202063}{\url{hep-ph/0202063}}}.

\bibitem[Giunti(2002)]{Giunti:2002xg}
C.~Giunti,
\newblock ``{Neutrino wave packets in quantum field theory}'',
\newblock {\em JHEP}, 11, 017, 2002,
\newblock
  \texttt{\href{http://arxiv.org/abs/hep-ph/0205014}{\url{hep-ph/0205014}}}.

\bibitem[Giunti(2004{\natexlab{a}})]{Giunti:2003ax}
C.~Giunti,
\newblock ``{Coherence and Wave Packets in Neutrino Oscillations}'',
\newblock {\em Found. Phys. Lett.}, 17, 103--124, 2004{\natexlab{a}},
\newblock
  \texttt{\href{http://arxiv.org/abs/hep-ph/0302026}{\url{hep-ph/0302026}}},
\newblock URL:
  \href{http://journals.kluweronline.com/article.asp?PIPS=486121}{\url{http://journals.kluweronline.com/article.asp?PIPS=486121}}.

\bibitem[Burkhardt et~al.(2003)Burkhardt, Lowe, Stephenson, and
  Goldman]{Burkhardt:hep-ph/0302084}
H.~Burkhardt, J.~Lowe, G.~J. Stephenson, and T.~Goldman,
\newblock ``{The wavelength of neutrino and neutral kaon oscillations}'',
\newblock {\em Phys. Lett.}, B566, 137, 2003,
\newblock
  \texttt{\href{http://arxiv.org/abs/hep-ph/0302084}{\url{hep-ph/0302084}}}.

\bibitem[Giunti(2003{\natexlab{b}})]{physics/0305122}
C.~Giunti,
\newblock ``{Lorentz Invariance of Neutrino Oscillations}'',
\newblock 2003{\natexlab{b}},
\newblock
  \texttt{\href{http://arxiv.org/abs/physics/0305122}{\url{physics/0305122}}}.

\bibitem[Giunti(2004{\natexlab{b}})]{hep-ph/0402217}
C.~Giunti,
\newblock ``{Flavor Neutrinos States}'',
\newblock 2004{\natexlab{b}},
\newblock
  \texttt{\href{http://arxiv.org/abs/hep-ph/0402217}{\url{hep-ph/0402217}}}.

\bibitem[Bilenky and Pontecorvo(1978)]{Bilenky:1978nj}
S.~M. Bilenky and B.~Pontecorvo,
\newblock ``{Lepton mixing and neutrino oscillations}'',
\newblock {\em Phys. Rept.}, 41, 225, 1978.

\bibitem[Bilenky and Petcov(1987)]{Bilenky:1987ty}
S.~M. Bilenky and S.~T. Petcov,
\newblock ``{Massive neutrinos and neutrino oscillations}'',
\newblock {\em Rev. Mod. Phys.}, 59, 671, 1987.

\bibitem[Boehm and Vogel(1992)]{Boehm:1992nn}
F.~Boehm and P.~Vogel,
\newblock {\em {Physics of massive neutrinos}},
\newblock Cambridge University Press, 1992.

\bibitem[Kim and Pevsner(1993)]{CWKim-book}
C.~W. Kim and A.~Pevsner,
\newblock {\em {Neutrinos in physics and astrophysics}},
\newblock Harwood Academic Press, Chur, Switzerland, 1993,
\newblock Contemporary Concepts in Physics, Vol. 8.

\bibitem[Zralek(1998)]{Zralek-oscillations-98}
M.~Zralek,
\newblock ``{From kaons to neutrinos: Quantum mechanics of particle
  oscillations}'',
\newblock {\em Acta Phys. Polon.}, B29, 3925, 1998,
\newblock
  \texttt{\href{http://arxiv.org/abs/hep-ph/9810543}{\url{hep-ph/9810543}}}.

\bibitem[Bilenky and Giunti(2001)]{Bilenkii:2001yh}
S.~M. Bilenky and C.~Giunti,
\newblock ``{Lepton numbers in the framework of neutrino mixing}'',
\newblock {\em Int. J. Mod. Phys.}, A16, 3931--3949, 2001,
\newblock
  \texttt{\href{http://arxiv.org/abs/hep-ph/0102320}{\url{hep-ph/0102320}}}.

\bibitem[Beuthe(2003)]{Beuthe:2001rc}
M.~Beuthe,
\newblock ``{Oscillations of neutrinos and mesons in quantum field theory}'',
\newblock {\em Phys. Rept.}, 375, 105--218, 2003,
\newblock
  \texttt{\href{http://arxiv.org/abs/hep-ph/0109119}{\url{hep-ph/0109119}}}.

\bibitem[Dolgov(2002)]{Dolgov:2002wy}
A.~D. Dolgov,
\newblock ``{Neutrinos in cosmology}'',
\newblock {\em Phys. Rept.}, 370, 333--535, 2002,
\newblock
  \texttt{\href{http://arxiv.org/abs/hep-ph/0202122}{\url{hep-ph/0202122}}}.

\bibitem[Giunti and Laveder()]{Neutrino-Unbound}
C.~Giunti and M.~Laveder,
\newblock ``{Neutrino Unbound}'',
\newblock \url{http://www.nu.to.infn.it}.

\bibitem[Bilenky et~al.(1999)Bilenky, Giunti, and Grimus]{BGG-review-98}
S.~M. Bilenky, C.~Giunti, and W.~Grimus,
\newblock ``{Phenomenology of neutrino oscillations}'',
\newblock {\em Prog. Part. Nucl. Phys.}, 43, 1, 1999,
\newblock
  \texttt{\href{http://arxiv.org/abs/hep-ph/9812360}{\url{hep-ph/9812360}}}.

\bibitem[Gonzalez-Garcia and Nir(2003)]{Gonzalez-Garcia:2002dz}
M.~Gonzalez-Garcia and Y.~Nir,
\newblock ``{Neutrino Masses and Mixing: Evidence and Implications}'',
\newblock {\em Rev. Mod. Phys.}, 75, 345--402, 2003,
\newblock
  \texttt{\href{http://arxiv.org/abs/hep-ph/0202058}{\url{hep-ph/0202058}}}.

\bibitem[Giunti and Laveder(2003)]{hep-ph/0310238}
C.~Giunti and M.~Laveder,
\newblock ``{Neutrino Mixing}'',
\newblock 2003,
\newblock
  \texttt{\href{http://arxiv.org/abs/hep-ph/0310238}{\url{hep-ph/0310238}}},
\newblock In 'Developments in Quantum Physics - 2004', edited by F. Columbus
  and V. Krasnoholovets, Nova Science Publishers, Inc. URL:
  \href{http://www.novapublishers.com/detailed_search.asp?id=1-59454-003-9}{\url{http://www.novapublishers.com/detailed_search.asp?id=1-59454-003-9}}.

\bibitem[Maltoni et~al.(2004)Maltoni, Schwetz, Tortola, and
  Valle]{hep-ph/0405172}
M.~Maltoni, T.~Schwetz, M.~Tortola, and J.~Valle,
\newblock ``{Status of global fits to neutrino oscillations}'',
\newblock 2004,
\newblock
  \texttt{\href{http://arxiv.org/abs/hep-ph/0405172}{\url{hep-ph/0405172}}}.

\bibitem[Bilenky et~al.(2003)Bilenky, Giunti, Grifols, and
  Masso]{Bilenky:2002aw}
S.~M. Bilenky, C.~Giunti, J.~A. Grifols, and E.~Masso,
\newblock ``{Absolute values of neutrino masses: Status and prospects}'',
\newblock {\em Phys. Rept.}, 379, 69--148, 2003,
\newblock
  \texttt{\href{http://arxiv.org/abs/hep-ph/0211462}{\url{hep-ph/0211462}}}.

\bibitem[Shrock(1980)]{Shrock:1980vy}
R.~E. Shrock,
\newblock ``{New tests for, and bounds on, neutrino masses and lepton
  mixing}'',
\newblock {\em Phys. Lett.}, B96, 159, 1980.

\bibitem[McKellar(1980)]{McKellar:1980cn}
B.~H.~J. McKellar,
\newblock ``{The influence of mixing of finite mass neutrinos on beta decay
  spectra}'',
\newblock {\em Phys. Lett.}, B97, 93, 1980.

\bibitem[Kobzarev et~al.(1980)Kobzarev, Martemyanov, Okun, and
  Shchepkin]{Kobzarev:1980nk}
I.~Y. Kobzarev, B.~V. Martemyanov, L.~B. Okun, and M.~G. Shchepkin,
\newblock ``{The phenomenology of neutrino oscillations}'',
\newblock {\em Sov. J. Nucl. Phys.}, 32, 823, 1980.

\bibitem[Shrock(1981{\natexlab{a}})]{Shrock:1981ct}
R.~E. Shrock,
\newblock ``{General theory of weak leptonic and semileptonic decays. 1.
  Leptonic pseudoscalar meson decays, with associated tests for, and bounds on,
  neutrino masses and lepton mixing}'',
\newblock {\em Phys. Rev.}, D24, 1232, 1981{\natexlab{a}}.

\bibitem[Shrock(1981{\natexlab{b}})]{Shrock:1981wq}
R.~E. Shrock,
\newblock ``{General theory of weak processes involving neutrinos. 2. Pure
  leptonic decays}'',
\newblock {\em Phys. Rev.}, D24, 1275, 1981{\natexlab{b}}.

\bibitem[Winter(1981)]{Winter:1981kj}
R.~G. Winter,
\newblock ``{Neutrino oscillation kinematics}'',
\newblock {\em Lett. Nuovo Cim.}, 30, 101--104, 1981.

\bibitem[Giunti(2003{\natexlab{c}})]{hep-ph/0311241}
C.~Giunti,
\newblock ``{Theory of Neutrino Oscillations}'',
\newblock 2003{\natexlab{c}},
\newblock
  \texttt{\href{http://arxiv.org/abs/hep-ph/0311241}{\url{hep-ph/0311241}}},
\newblock IFAE 2003, Lecce, 23-26 April 2003. URL:
  \href{http://www.le.infn.it/ifae/PDF/Giunti.pdf}{\url{http://www.le.infn.it/ifae/PDF/Giunti.pdf}}.

\bibitem[Giunti(2004{\natexlab{c}})]{hep-ph/0401244}
C.~Giunti,
\newblock ``{Theory of Neutrino Oscillations}'',
\newblock 2004{\natexlab{c}},
\newblock
  \texttt{\href{http://arxiv.org/abs/hep-ph/0401244}{\url{hep-ph/0401244}}},
\newblock 11th Lomonosov Conference on Elementary Particle Physics, 21-27
  August 2003, Moscow State University, Moscow, Russia.

\bibitem[Sassaroli et~al.(1995)Sassaroli, Srivastava, and
  Widom]{hep-ph/9509261}
E.~Sassaroli, Y.~N. Srivastava, and A.~Widom,
\newblock ``{Charged Lepton Oscillations}'',
\newblock {\em Z. Phys. C66}, pages 601--605, 1995,
\newblock
  \texttt{\href{http://arxiv.org/abs/hep-ph/9509261}{\url{hep-ph/9509261}}}.

\bibitem[Srivastava and Widom(1997)]{hep-ph/9707268}
Y.~N. Srivastava and A.~Widom,
\newblock ``{Of course muons can oscillate}'',
\newblock 1997,
\newblock
  \texttt{\href{http://arxiv.org/abs/hep-ph/9707268}{\url{hep-ph/9707268}}}.

\bibitem[Srivastava et~al.(1998{\natexlab{a}})Srivastava, Widom, and
  Sassaroli]{Srivastava:1998gi}
Y.~Srivastava, A.~Widom, and E.~Sassaroli,
\newblock ``{Charged lepton and neutrino oscillations}'',
\newblock {\em Eur. Phys. J.}, C2, 769, 1998{\natexlab{a}}.

\bibitem[Srivastava et~al.(1995{\natexlab{a}})Srivastava, Widom, and
  Sassaroli]{Srivastava:1995ws}
Y.~N. Srivastava, A.~Widom, and E.~Sassaroli,
\newblock ``{Lambda oscillations}'',
\newblock {\em Phys. Lett.}, B344, 436--440, 1995{\natexlab{a}}.

\bibitem[Widom and Srivastava(1996{\natexlab{a}})]{hep-ph/9605399}
A.~Widom and Y.~N. Srivastava,
\newblock ``{Lambda oscillations and the conservation laws}'',
\newblock {\em Phys. Lett. B344}, pages 436--440, 1996{\natexlab{a}},
\newblock
  \texttt{\href{http://arxiv.org/abs/hep-ph/9605399}{\url{hep-ph/9605399}}}.

\bibitem[Lowe et~al.(1996)]{Lowe:hep-ph/9605234}
J.~Lowe et~al.,
\newblock ``{No $\Lambda$ oscillations}'',
\newblock {\em Phys. Lett.}, B384, 288--292, 1996,
\newblock
  \texttt{\href{http://arxiv.org/abs/hep-ph/9605234}{\url{hep-ph/9605234}}}.

\bibitem[Srivastava et~al.(1998{\natexlab{b}})Srivastava, Palit, Widom, and
  Sassaroli]{Srivastava:hep-ph/9807543}
Y.~N. Srivastava, S.~Palit, A.~Widom, and E.~Sassaroli,
\newblock ``{Neutrino mass difference induced oscillations in observed muon
  decays}'',
\newblock 1998{\natexlab{b}},
\newblock
  \texttt{\href{http://arxiv.org/abs/Srivastava:hep-ph/9807543}{\url{Srivastava:hep-ph/9807543}}}.

\bibitem[Field(2003)]{Field:hep-ph/0211199}
J.~H. Field,
\newblock ``{A covariant path amplitude description of flavour oscillations:
  The Gribov-Pontecorvo phase for neutrino vacuum propagation is right}'',
\newblock {\em Eur. Phys. J.}, C30, 305--325, 2003,
\newblock
  \texttt{\href{http://arxiv.org/abs/hep-ph/0211199}{\url{hep-ph/0211199}}}.

\bibitem[Grossman and Lipkin(1997)]{hep-ph/9607201}
Y.~Grossman and H.~J. Lipkin,
\newblock ``{Flavor oscillations from a spatially localized source: A simple
  general treatment}'',
\newblock {\em Phys. Rev.}, D55, 2760--2767, 1997,
\newblock
  \texttt{\href{http://arxiv.org/abs/hep-ph/9607201}{\url{hep-ph/9607201}}}.

\bibitem[Leo et~al.(2000{\natexlab{a}})Leo, Ducati, and
  Rotelli]{hep-ph/9906460}
S.~D. Leo, G.~Ducati, and P.~Rotelli,
\newblock ``{Comments upon the mass oscillation formulas}'',
\newblock {\em Mod. Phys. Lett.}, A15, 2057--2068, 2000{\natexlab{a}},
\newblock
  \texttt{\href{http://arxiv.org/abs/hep-ph/9906460}{\url{hep-ph/9906460}}}.

\bibitem[De~Leo et~al.(2002)De~Leo, Nishi, and Rotelli]{DeLeo:hep-ph/0208086}
S.~De~Leo, C.~C. Nishi, and P.~P. Rotelli,
\newblock ``{Quantum oscillation phenomena}'',
\newblock 2002,
\newblock
  \texttt{\href{http://arxiv.org/abs/hep-ph/0208086}{\url{hep-ph/0208086}}}.

\bibitem[Lipkin(1999)]{Lipkin:hep-ph/9901399}
H.~J. Lipkin,
\newblock ``{Quantum mechanics of neutrino oscillations: Hand waving for
  pedestrians}'',
\newblock 1999,
\newblock
  \texttt{\href{http://arxiv.org/abs/hep-ph/9901399}{\url{hep-ph/9901399}}}.

\bibitem[Srivastava et~al.(1995{\natexlab{b}})Srivastava, Widom, and
  Sassaroli]{Srivastava:1995bg}
Y.~Srivastava, A.~Widom, and E.~Sassaroli,
\newblock ``{Spatial correlations in two neutral kaon decays}'',
\newblock {\em Z. Phys.}, C66, 601--605, 1995{\natexlab{b}}.

\bibitem[Widom and Srivastava(1996{\natexlab{b}})]{Widom:hep-ph/9605399}
A.~Widom and Y.~N. Srivastava,
\newblock ``{Lambda oscillations and the conservation laws}'',
\newblock 1996{\natexlab{b}},
\newblock
  \texttt{\href{http://arxiv.org/abs/hep-ph/9605399}{\url{hep-ph/9605399}}}.

\bibitem[Lipkin(1995)]{hep-ph/9501269}
H.~J. Lipkin,
\newblock ``{Theories of nonexperiments in coherent decays of neutral
  mesons}'',
\newblock {\em Phys. Lett.}, B348, 604--608, 1995,
\newblock
  \texttt{\href{http://arxiv.org/abs/hep-ph/9501269}{\url{hep-ph/9501269}}}.

\bibitem[Nieto(1996)]{Nieto:hep-ph/9509370}
M.~M. Nieto,
\newblock ``{Quantum Interference: From Kaons to Neutrinos (with Quantum Beats
  in between)}'',
\newblock {\em Hyperfine Interact.}, 100, 193, 1996,
\newblock
  \texttt{\href{http://arxiv.org/abs/hep-ph/9509370}{\url{hep-ph/9509370}}}.

\bibitem[Kayser and Stodolsky(1995)]{Kayser:1995bw}
B.~Kayser and L.~Stodolsky,
\newblock ``{EPR experiments without 'collapse of the wave function'}'',
\newblock {\em Phys. Lett.}, B359, 343--350, 1995.

\bibitem[Kayser(1997{\natexlab{a}})]{Kayser:1997fr}
B.~Kayser,
\newblock ``{The frequency of neutral meson and neutrino oscillation}'',
\newblock 1997{\natexlab{a}},
\newblock SLAC-PUB-7123. URL:
  \href{http://www.slac.stanford.edu/pubs/slacpubs/7000/slac-pub-7123.html}{\url{http://www.slac.stanford.edu/pubs/slacpubs/7000/slac-pub-7123.html}}.

\bibitem[Kayser(1997{\natexlab{b}})]{Kayser:hep-ph/9702327}
B.~Kayser,
\newblock ``{CP violation, mixing, and quantum mechanics}'',
\newblock 1997{\natexlab{b}},
\newblock
  \texttt{\href{http://arxiv.org/abs/hep-ph/9702327}{\url{hep-ph/9702327}}},
\newblock 28th International Conference on High-energy Physics (ICHEP 96),
  Warsaw, Poland, 25-31 Jul 1996.

\bibitem[Leo et~al.(2000{\natexlab{b}})Leo, Ducati, and Rotelli]{Rotelli-99}
S.~D. Leo, G.~Ducati, and P.~Rotelli,
\newblock ``{Comments upon the mass oscillation formulas}'',
\newblock {\em Mod. Phys. Lett.}, A15, 2057--2068, 2000{\natexlab{b}},
\newblock
  \texttt{\href{http://arxiv.org/abs/hep-ph/9906460}{\url{hep-ph/9906460}}}.

\bibitem[Blasone and Vitiello(1995)]{Blasone:1995zc}
M.~Blasone and G.~Vitiello,
\newblock ``{Quantum field theory of fermion mixing}'',
\newblock {\em Ann. Phys.}, 244, 283--311, 1995,
\newblock
  \texttt{\href{http://arxiv.org/abs/hep-ph/9501263}{\url{hep-ph/9501263}}}.

\bibitem[Blasone et~al.(1999)Blasone, Henning, and Vitiello]{Blasone:1998hf}
M.~Blasone, P.~A. Henning, and G.~Vitiello,
\newblock ``{The exact formula for neutrino oscillations}'',
\newblock {\em Phys. Lett.}, B451, 140, 1999,
\newblock
  \texttt{\href{http://arxiv.org/abs/hep-th/9803157}{\url{hep-th/9803157}}}.

\bibitem[Fujii et~al.(1999)Fujii, Habe, and Yabuki]{Fujii:1998xa}
K.~Fujii, C.~Habe, and T.~Yabuki,
\newblock ``{Note on the field theory of neutrino mixing}'',
\newblock {\em Phys. Rev.}, D59, 113003, 1999,
\newblock
  \texttt{\href{http://arxiv.org/abs/hep-ph/9807266}{\url{hep-ph/9807266}}}.

\bibitem[Blasone and Vitiello(1999)]{hep-ph/9907382}
M.~Blasone and G.~Vitiello,
\newblock ``{Remarks on the neutrino oscillation formula}'',
\newblock {\em Phys. Rev.}, D60, 111302, 1999,
\newblock
  \texttt{\href{http://arxiv.org/abs/hep-ph/9907382}{\url{hep-ph/9907382}}}.

\bibitem[Fujii et~al.(2001)Fujii, Habe, and Yabuki]{Fujii:2001zv}
K.~Fujii, C.~Habe, and T.~Yabuki,
\newblock ``{Remarks on flavor-neutrino propagators and oscillation
  formulae}'',
\newblock {\em Phys. Rev.}, D64, 013011, 2001,
\newblock
  \texttt{\href{http://arxiv.org/abs/hep-ph/0102001}{\url{hep-ph/0102001}}}.

\bibitem[Blasone et~al.(2002)Blasone, Capolupo, and Vitiello]{Blasone:2002jv}
M.~Blasone, A.~Capolupo, and G.~Vitiello,
\newblock ``{Quantum field theory of three flavor neutrino mixing and
  oscillations with CP violation}'',
\newblock {\em Phys. Rev.}, D66, 025033, 2002,
\newblock
  \texttt{\href{http://arxiv.org/abs/hep-th/0204184}{\url{hep-th/0204184}}}.

\bibitem[Blasone et~al.(2003)Blasone, Pacheco, and Tseung]{Blasone:2002wp}
M.~Blasone, P.~P. Pacheco, and H.~W.~C. Tseung,
\newblock ``{Neutrino oscillations from relativistic flavor currents}'',
\newblock {\em Phys. Rev.}, D67, 073011, 2003,
\newblock
  \texttt{\href{http://arxiv.org/abs/hep-ph/0212402}{\url{hep-ph/0212402}}}.

\bibitem[Giunti(2003{\natexlab{d}})]{hep-ph/0312256}
C.~Giunti,
\newblock ``{Fock States of Flavor Neutrinos are Unphysical}'',
\newblock 2003{\natexlab{d}},
\newblock
  \texttt{\href{http://arxiv.org/abs/hep-ph/0312256}{\url{hep-ph/0312256}}}.

\end{thebibliography}
\end{document}